\documentclass[aps,prx,reprint,superscriptaddress]{revtex4-2}
\usepackage[T1]{fontenc}
\usepackage[latin9]{inputenc}
\usepackage{color}
\usepackage{babel}
\usepackage{amsmath}
\usepackage{amssymb}
\usepackage{graphicx}
\usepackage{placeins}
\usepackage[unicode=true,pdfusetitle,
 bookmarks=true,bookmarksnumbered=false,bookmarksopen=false,
 breaklinks=false,pdfborder={0 0 1},backref=false,colorlinks=true]
 {hyperref}
\hypersetup{
 allcolors=blue}

\makeatletter

\providecommand{\tabularnewline}{\\}

\usepackage[]{ulem}
\usepackage{txfonts}
\usepackage{braket}
\newcommand{\hide}[1]{}
\newcommand{\sect}[1]{\vspace{0.2cm}\noindent \textbf{\large{#1}}}
\newcommand{\subsect}[1]{\noindent \textbf{#1}. }

\makeatother

\begin{document}
\title{Theory of rare-earth Kramers magnets on a Shastry-Sutherland lattice:
\\
dimer phases in the presence of strong spin-orbit coupling}
\author{Changle Liu}
\thanks{These authors contributed equally to the work.}
\affiliation{School of Physics and Mechatronic Engineering, Guizhou Minzu University, Guiyang 550025, China}
\author{Guijing Duan}
\thanks{These authors contributed equally to the work.}
\affiliation{School of Physics and Beijing Key Laboratory of Opto-electronic Functional
Materials \& Micro-nano Devices, Renmin University of China, Beijing
100872, China}
\author{Rong Yu}
\email{rong.yu@ruc.edu.cn}
\affiliation{School of Physics and Beijing Key Laboratory of Opto-electronic Functional
Materials \& Micro-nano Devices, Renmin University of China, Beijing
100872, China}
\affiliation{Key Laboratory of Quantum State Construction and Manipulation (Ministry
of Education), Renmin University of China, Beijing, 100872, China}

\begin{abstract}
	Shastry-Sutherland magnet is a typical frustrated spin system hosting rich phases. While the Heisenberg limit has been extensively studied, the role of spin-orbit coupling is not well explored. Motivated by newly discovered rare-earth Shastry-Sutherland magnets, we construct a generic effective-spin model that describes the interactions between Kramers doublet local moments on a Shastry-Sutherland lattice. Due to the strong spin-orbit coupling, the model takes the form of extended XYZ interactions on both intra- and inter-dimer bonds. We show that, in addition to the conventional ``singlet'' dimer phase, strong spin-orbit coupling can stabilize peculiar ``triplet'' dimer phases. These ``triplet'' dimer phases, though fully gapped, respond immediately to magnetic fields and evolve smoothly into the fully polarized phase. We present that the recently discovered Shastry-Sutherland magnet Yb$_{2}$Be$_{2}$GeO$_{7}$ belongs to the ``triplet'' dimer phase, and discuss the implication of our results to a broad class of quantum magnets in general.
\end{abstract}
\date{\today}

\maketitle

%\sect{Introduction}

Quantum magnetism on the Shastry-Sutherland (SS)
lattice has attracted significant attention over decades due to its simple 
lattice structure yet resultant rich physical behaviors \cite{shastry1981exact,miyahara1999exact,kageyama1999exact,koga2000quantum, mcclarty2017topological,song2020abnormal,guo2020quantum,cui2023proximate,nomura2023unveiling}.
As one of the simplest geometrically frustrated systems in two spatial
dimensions, the SS model, the $S=1/2$ Heisenberg antiferromagnetic
model defined on the SS lattice, was initially proposed because of
the exact solvability of its dimer singlet ground state within a certain
parameter regime \cite{shastry1981exact,kageyama1999exact}. Meanwhile,
this dimer state also exhibits highly localized triplet excitations,
and is responsible for the emergence of various magnetization plateau 
phases under an external magnetic field 
\cite{miyahara2003theory,zayed20174,muller2000exact,onizuka20001, 
takigawa200418,levy2008field,jaime2012magnetostriction}.
The strong geometrical frustration of the SS model is also clearly
manifested in its condensed matter realization, SrCu$_{2}$(BO$_{3}$)$_{2}$
\cite{miyahara1999exact,koga2000quantum,levy2008field, 
jaime2012magnetostriction,guo2020quantum,nomura2023unveiling,cui2023proximate, 
wang2023plaquette,jimenez2021quantum,cui2024two,NPJQM2021,fan2024}.
In this compound, the competition between the inter- and intra-dimer
interactions, $J$ and $J^{\prime}$, leads to a dimerized ground
state at ambient pressure. Nearly flat-band excitations have been
observed in inelastic neutron scattering \cite{haravifard2012neutron,zayed2014correlated},
and various magnetization plateau phases predicted in theory are 
observed in the in-field phase diagram, supporting 
the stabilization of the dimer singlet ground state.
Moreover, when applying a hydrostatic
pressure, the ratio between the inter- and intra-dimer couplings $J/J^{\prime}$
increases, and the ground state evolves to the plaquette and N\'eel
antiferromagnetic (AFM) phases subsequently \cite{zayed20174,lee2019signatures,guo2020quantum,jimenez2021quantum}.
The transition from the plaquette to the AFM phase is theoretically
proposed to be associated with deconfined quantum criticality, and
has attracted much experimental and numerical interest in recent years
\cite{zayed20174,cui2023proximate,lee2019signatures,xi2023plaquette, 
corboz2013tensor,yang2022quantum,chen2024spin}.

Recently, research on quantum magnets has been extended to systems
with strong spin-orbit coupling (SOC), pushing the boundaries of condensed
matter. %\cite{rousochatzakis2024beyond,takayama2015hyperhoneycomb, 
%baek2017evidence,zhang2022anisotropic,smith2023quantum,hallas2018experimental, 
%liu2023ba9re2,gao2024spin,li2015rare,ma2018spin,kimchi2018valence, 
%xie2023complete,ashtar2019synthesis,shen2016evidence,paddison2017continuous, 
%baek2017evidence,kim2015kitaev,rau2019frustrated,li2018effect, 
%bordelon2019field,ishii2021magnetic} 
The strong SOC can give rise to highly anisotropic interactions
that frustrate local moments, resulting in highly correlated states
with lack of long-range order known as quantum spin liquids (QSL)
%\cite{shen2016evidence,paddison2017continuous,baek2017evidence},
\cite{balents2010spin,savary2016quantum,zhou2017quantum,broholm2020quantum},
characterized by properties such as fractionalized excitations, emergent
gauge structures, and long-range quantum entanglement. Driven by such
possibilities, there have been intense experimental and theoretical
explorations in the past decade on relevant systems such as honeycomb
Kitaev systems \cite{kim2015kitaev,baek2017evidence,takayama2015hyperhoneycomb, 
rousochatzakis2024beyond,zhang2022anisotropic,liu2023ba9re2, 
jang2019antiferro,jang2020computational}, pyrochlore spin ice 
\cite{rau2019frustrated,smith2023quantum,hallas2018experimental},
and large classes of triangular lattice materials 
\cite{li2018effect,paddison2017continuous,shen2016evidence,bordelon2019field, 
li2015rare,paddison2017continuous,gao2024spin,li2015rare,ma2018spin, 
kimchi2018valence,xie2023complete,ashtar2019synthesis}. Moreover, the interplay 
between strong SOC and the surrounding crystalline
electric field (CEF) environment can give rise to complex forms of
multipolar ordering~\cite{NPJQM2023}. Aside from magnetic insulators, the multipolar
orders have intimate connections to heavy fermion systems, and are
responsible for exotic phenomena such as non-Fermi liquid behaviors,
unconventional superconductivity, and intricate magnetic or electric
responses therein.

Currently, there have been explodingly increasing reports on discoveries
of rare-earth SS materials 
\cite{li2024spinons,brassington2024magnetic,brassington2024synthesis, 
liu2024distinct,ishii2021magnetic,yadav2024observation,YBGO, 
ashtar2021structure,pula2024candidate,nonkramers,EBGO,brassington2025novel,pula2025ground, EuMgSiO}.
For these systems, two basic questions related to SOC naturally arise:
The first one is whether the conclusions holding for the Heisenberg
system, such as the exact solvability and highly localized triplet
excitations of the dimer phase, would still be  valid in the presence
of strong SOC. The second one is whether exotic phases or phenomena
could be stabilized in the phase diagram by the introduction of SOC. 
Addressing the above questions requires collective efforts from both experiment
and theory. On the theory side, it is highly appreciated if we can
have understanding on the microscopic details of relevant materials, including
the structure of local moments, as well as the interactions among them.
Such knowledge, bridging the theoretical models and experimental observables,
would be particularly helpful in analyzing the experimental behaviors
of relevant systems.

In this work, we develop a theory for rare-earth SS magnets, focusing
on the case of Kramers ions. We establish an effective model that
describes the interactions among rare-earth local moments. Remarkably,
we find that strong SOC can stabilize peculiar dimerized phases in
the phase diagram, which we refer to as ``triplet'' dimer phases.
These phases, as previously unreported in strong SOC magnets, exhibit
intriguing magnetic responses: while they possess a finite 
excitation gap, 
they can be magnetized under an infinitesimal
external magnetic field. Moreover, these phases are adiabatically
connected to the high-field polarized phase without any phase transition.
These unusual properties distinguish them from the conventional 
dimer singlet state. We have performed a comprehensive exploration on
the experimental signatures of triplet dimer phases in various measurements,
and find that the triplet dimer scenario offers a natural explanation
for the puzzling observations on a number of 
dimerized magnets with strong SOC, such as
Yb$_{2}$Be$_{2}$GeO$_{7}$. Moreover, we find that the concept of 
``triplet'' dimers can be extended to a broader context
within quantum magnetism.

\begin{figure}
	\includegraphics[width=\columnwidth]{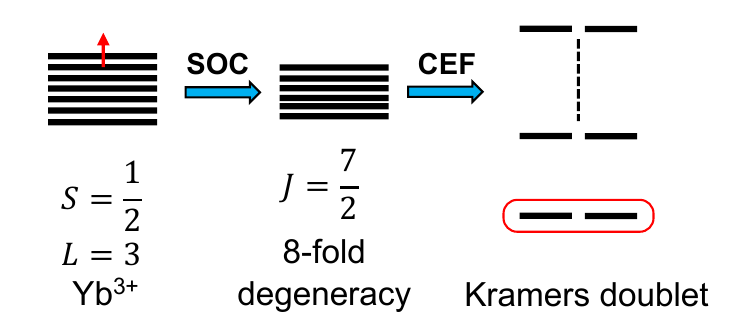} 
	\caption{\textbf{Crystal field scheme of Yb$^{3+}$ in the rare-earth SS magnet
			Yb$_{2}$Be$_{2}$GeO$_{7}$}. The red arrow represents one hole in
		the 4$f$ shell. Under strong SOC and CEF the electronic levels split
		into a set of Kramers doublets. At low temperatures, the magnetism
		is %contributed
		dominated by the lowest Kramers doublet which is well separated to
		other Kramers doublets.}
	\label{fig:CEF}
\end{figure}

\sect{Results}

\subsect{Local moment structure}
We consider the crystal structure of a class of SS
magnets, $RE_{2}$Be$_{2}$\textit{X}O$_{7}$, where \textit{X}=Si,
Ge, and $RE^{3+}$ (\textit{RE}=Yb, Er, Dy, \textit{etc.}) are rare-earth
Kramers ions that form a perfect SS lattice \cite{ashtar2021structure,brassington2024synthesis}.
Without loss of generality, here we take the Yb$_{2}$Be$_{2}$\textit{X}O$_{7}$
compound \cite{pula2024candidate,YBGO,brassington2025novel} 
as an example,
and general results are applicable to all Kramers ions other than
Gd$^{3+}$ (given that Gd$^{3+}$ does not carry orbital angular momentum
and hence does not have strong SOC). The crystal field scheme of Yb$^{3+}$
is shown in Fig. \ref{fig:CEF}. The magnetism of Yb$^{3+}$ ion is
contributed by the partially filled electrons within the $4f$ shell.
Each Yb$^{3+}$ ion carries an orbital angular momentum $L=3$ and spin
angular momentum $S=1/2$. The strong atomic %spin-orbit coupling 
SOC entangles
the spin and orbital degrees of freedom, leading to a total angular
momentum $\mathcal{J}=7/2$ at low energies. The eight-fold $\mathcal{J}=7/2$
electronic levels are further lifted by the CEF, and split into a
series of Kramers doublets protected by the time-reversal symmetry. When
the temperature is lower than the crystal field splitting between
different Kramers doublets, the magnetism is dominated by the lowest
Kramers doublet.

\begin{figure*}
	\includegraphics[width=1.5\columnwidth]{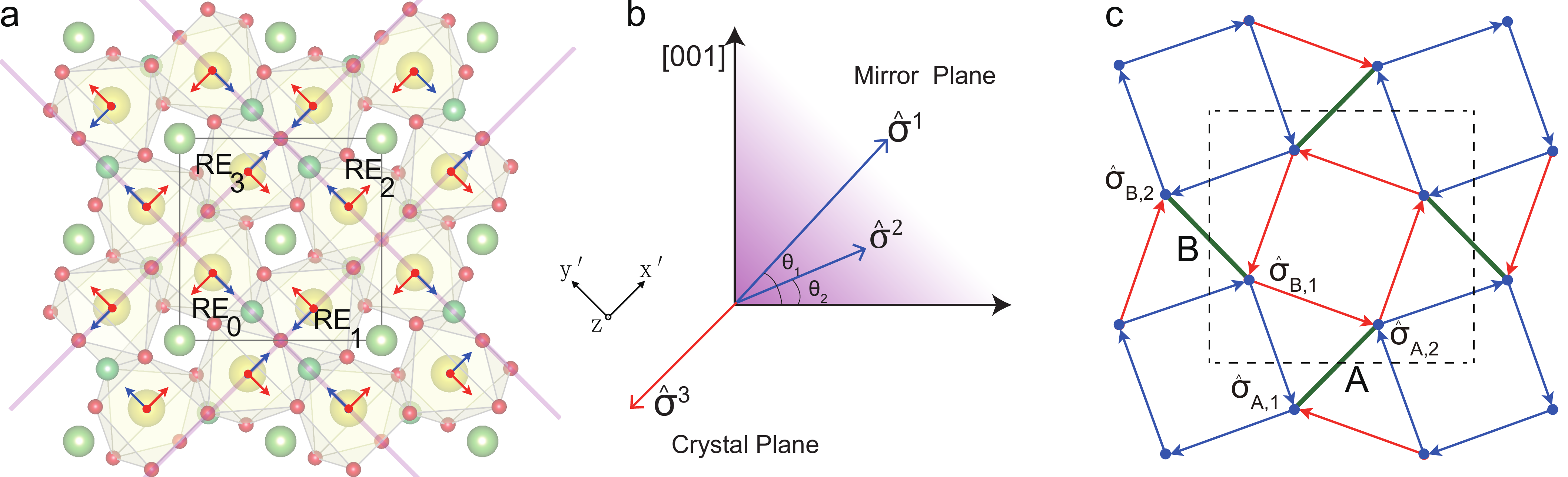}
	\caption{\textbf{Crystal structure and dipole axes of SS magnets with Kramers
			local moments}. \textbf{a} Crystal structure of Yb$_{2}$Be$_{2}$GeO$_{7}$.
		Yb$^{3+}$ and O$^{2-}$ are denoted by yellow and red balls, respectively.
		Mirror planes of Yb$^{3+}$ are denoted by purple lines. The in-plane
		directions of $\hat{\sigma}^{1}$ and $\hat{\sigma}^{2}$ dipole axes
		are denoted by the blue arrows, while that of $\hat{\sigma}^{3}$
		are indicated by the red arrows. \textbf{b} Illustration of dipole
		axes. The $\hat{\sigma}^{1}$ and $\hat{\sigma}^{2}$ components are
		within the mirror plane, while the $\hat{\sigma}^{3}$ component is
		perpendicular to the mirror plane. $\hat{\sigma}^{1}$ and $\hat{\sigma}^{2}$
		are generally non-orthogonal. \textbf{c} Illustration of the SS lattice.
		Each unit cell contains two dimer bonds denoted by the thick dark
		green lines and marked by $A$, $B$, respectively. The NN inter-dimer
		bonds can also be divided into two groups, characterized by $\eta_{ij}=\pm1$
		in Eq. (\ref{eq:inter-dimer}), and are marked red and blue, respectively,
		with the directions $i\rightarrow j$ indicated by arrows.}
	\label{fig:crystal}
\end{figure*}

The structure of the Kramers local moments can be analyzed 
based on symmetries
of the CEF levels. We define the $z$ direction to be perpendicular
to the plane of %local moments
the magnetic ions,
 and $x^{\prime}$ and $y^{\prime}$
are along the two orthogonal dimer directions (see Fig. \ref{fig:crystal}\textbf{a}).
As the four $RE^{3+}$ sublattices within the unit cell (labeled by
RE$_{i}$, $i=0,1,2,3$ as shown in Fig. \ref{fig:crystal}\textbf{a})
are connected by the four-fold roto-inversion operation $S_{4}$ about
the $z$-axis located at the cell center, we can start with the RE$_{1}$
sublattice, and the crystal fields of the other three sites are related
by the corresponding symmetry operations. Note that the point group
symmetry of each $RE^{3+}$ ion only contains a mirror reflection
$\sigma_{v}$ with the mirror plane parallel to the corresponding
dimer direction (mirror planes are indicated by pink solid lines in
Fig. \ref{fig:crystal}\textbf{a}). For the RE$_{1}$ sublattice,
we denote the lowest Kramers doublet as $|\psi_{\pm}\rangle$. Since
the Kramers ion has a half-integer total angular momentum, 
the eigenvalues of the mirror reflection $\sigma_{v}$ %become 
are constrained to be pure imaginary $\pm i$, and this is quite different 
from the non-Kramers case. Denoting $|\psi_{\pm}\rangle$
for the eigenstates of the mirror reflection symmetry, \textit{e.g.},
$\sigma_{v}|\psi_{\pm}\rangle=\pm i|\psi_{\pm}\rangle$, we can construct
an effective spin-1/2 operator by acting the Pauli matrices $\tau^{\alpha}$
onto the Kramers doublet: $\hat{\sigma}^{\alpha}\equiv\frac{1}{2}\sum_{\mu\nu}|\psi_{\mu}\rangle\tau_{\mu\nu}^{\alpha}\langle\psi_{\nu}|=\frac{1}{2}\psi\tau^{\alpha}\psi^{\dagger}$
where $\psi=\left(|\psi_{+}\rangle,|\psi_{-}\rangle\right)$ and $\alpha=1,2,3$.
The symmetry transformations of effective spin components under the
mirror reflection $\sigma_{v}$ and time-reversal $\Theta$ are as
follows:
\[
\sigma_{v}:\hat{\sigma}^{1}\rightarrow-\hat{\sigma}^{1},\hat{\sigma}^{2}\rightarrow-\hat{\sigma}^{2},\hat{\sigma}^{3}\rightarrow\hat{\sigma}^{3},
\]

\[
\Theta:\hat{\sigma}^{1}\rightarrow-\hat{\sigma}^{1},\hat{\sigma}^{2}\rightarrow-\hat{\sigma}^{2},\hat{\sigma}^{3}\rightarrow-\hat{\sigma}^{3}.
\]

As shown in the above relation, all the three effective spin components
are odd under the time-reversal and are thus magnetic dipoles. Moreover,
both $\hat{\sigma}^{1}$ and $\hat{\sigma}^{2}$ are odd under the
mirror reflection $\sigma_{v}$. Hence the corresponding dipole moments
lie within the $x^{\prime}z$ mirror plane (transform as $\mathcal{J}^{x^{\prime}}$
or $\mathcal{J}^{z}$ under crystalline symmetries), with specific
angles $\theta_{1}$ and $\theta_{2}$, respectively, to the crystal
plane. Meanwhile, $\hat{\sigma}^{3}$ is even under $\sigma_{v}$
that transforms as $\mathcal{J}^{y^{\prime}}$, hence its moment direction
is perpendicular to the mirror plane, as illustrated in Fig. \ref{fig:crystal}\textbf{b}.
More quantitatively, the dipolar character of local moments discussed
above can be clearly manifested by projecting the total angular momentum
$\boldsymbol{\mathcal{J}}$ onto the two-level subspace:

\begin{align}
\hat{\mathbf{j}} & \equiv\mathcal{P}\boldsymbol{\mathcal{J}}\mathcal{P}\nonumber \\
 & =A^{(1)}\hat{\sigma}^{1}\mathbf{n}^{(1)}+A^{(2)}\hat{\sigma}^{2}\mathbf{n}^{(2)}+A^{(3)}\hat{\sigma}^{3}\mathbf{n}^{(3)},
\end{align}
where $\mathcal{P}\equiv\psi\psi^{\dagger}=|\psi_{+}\rangle\langle\psi_{+}|+|\psi_{-}\rangle\langle\psi_{-}|$
is the projection operator onto the lowest doublet subspace, $A^{(\alpha)}$
describes the magnitude of the dipole moment $\hat{\sigma}^{\alpha}$,
and $\mathbf{n}^{(\alpha)}$ are unit vectors representing the directions
of dipoles $\hat{\sigma}^{\alpha}$ ($\alpha=1,2,3$). Note that $\mathbf{n}^{(3)}$
is completely fixed by %symmetries 
symmetry to be perpendicular to the mirror
plane. However, %the symmetry cannot fully determine 
the directions of $\mathbf{n}^{(1)}$ and $\mathbf{n}^{(2)}$ cannot be fully determined by 
symmetry: They are constrained
to lie within the mirror plane, but their precise directions 
depend on the exact wave functions of the crystal fields. As
a consequence, the dipole axes are in general non-orthogonal, \textit{i.e.},
$\mathbf{n}^{(1)}\cdot\mathbf{n}^{(2)}\neq0$, due to the low point
group symmetries of the rare-earth sites.

The above discussions are restricted to the RE$_{1}$ ion. As the
four RE$_{i}$ sublattices within the unit cell are connected by the
four-fold roto-inversion $S_{4}$, the directions of dipoles at different
sites $\mathbf{n}_{i}^{(\alpha)}$ are connected by the corresponding
four-fold counter-rotation about the $z$ axis. Therefore, for different
sublattices, $\mathbf{n}_{i}^{(1)}$ and $\mathbf{n}_{i}^{(2)}$ must
have the same out-of-plane component, while the directions of in-plane
components are illustrated by the blue arrows in Fig. \ref{fig:crystal}\textbf{a}.
Meanwhile, $\mathbf{n}_{i}^{(3)}$ are located within the crystal
plane with directions %as 
shown by the red arrows in Fig. \ref{fig:crystal}\textbf{a}.

\subsect{Effective Hamiltonian} %\label{subsec:Hamiltonian}
Here we derive a generic effective Hamiltonian based on symmetry analysis.
Due to the strongly localized nature of the 4$f$ electrons, it is
sufficient to consider only the intra-dimer and the nearest-neighbor
(NN) inter-dimer interactions, as well as the Zeeman coupling to the external magnetic field, where the total Hamiltonian
takes the form
\begin{equation}
	H=H_{J}+H_{J^{\prime}}+H_\textrm{Zeeman}.\label{eq:ham}
\end{equation}.

We begin with the intra-dimer interaction $H_{J^{\prime}}$. For each dimer bond, there
are two relevant mirror symmetries $\sigma_{v}$ and $\sigma_{v}^{\prime}$,
with the mirror planes located parallel and perpendicular to the bond
directions, respectively. Suppose the generic intra-dimer interactions
are of the following bilinear form: $H_{J'}=\sum_{\langle\langle 
ij\rangle\rangle\alpha\beta} 
J{}_{ij}^{\prime\alpha\beta}\hat{\sigma}_{i}^{\alpha}\hat{\sigma}_{j}^{\beta}.$
Since $\hat{\sigma}^{3}$ has different parity from $\hat{\sigma}^{1}$
and $\hat{\sigma}^{2}$ under the mirror $\sigma_{v}$, they are not allowed
to mix linearly, and hence  $J_{ij}^{\prime13}=J{}_{ij}^{\prime23} 
=J{}_{ij}^{\prime31}=J{}_{ij}^{\prime32}=0$.
Moreover, the $\sigma_{v}^{\prime}$ mirror reflection exchanges the
positions of the two sites of the dimer. This guarantees 
$J_{ij}^{\prime12}=J_{ij}^{\prime21}$.
As a consequence, the anti-symmetric Dzyaloshinskii-Moriya (DM) interaction
\cite{DM1958,DM1960} is disallowed, and the intra-dimer interaction
simply takes the form

\[
H_{J^{\prime}}=\sum_{\langle\langle ij\rangle\rangle}\left(\hat{\sigma}_{i}^{1},\hat{\sigma}_{i}^{2},\hat{\sigma}_{i}^{3}\right)\begin{pmatrix}J^{\prime11} & J^{\prime12}\\
J^{\prime12} & J^{\prime22}\\
 &  & J^{\prime33}
\end{pmatrix}\begin{pmatrix}\hat{\sigma}_{j}^{1}\\
\hat{\sigma}_{j}^{2}\\
\hat{\sigma}_{j}^{3}
\end{pmatrix}.
\]
The form of interaction can be further simplified if we perform a
global basis rotation on the CEF basis $|\psi_{\pm}\rangle$ at each
site, so that the cross-coupling term $J^{\prime12}$ can be eliminated.
The resulting interaction is an XYZ model 
\begin{equation}
H_{J^{\prime}}=\sum_{\langle\langle ij\rangle\rangle} 
J^{\prime11}\hat{\sigma}_{i}^{1}\hat{\sigma}_{j}^{1} 
+J^{\prime22}\hat{\sigma}_{i}^{2}\hat{\sigma}_{j}^{2} 
+J^{\prime33}\hat{\sigma}_{i}^{3}\hat{\sigma}_{j}^{3},
\end{equation}
similar to the case of non-Kramers systems~\cite{nonkramers}.
Note that although this system lacks inversion symmetry about the
dimer center, intra-dimer DM interaction is still forbidden. 
The absence of the intra-dimer DM interaction is closely related to the
local coordinate axes for strong SOC systems, and will be discussed in more 
detail in \textit{Discussion}.

Then we consider the NN inter-dimer interaction that takes the generic
form $H_{J}=\sum_{\langle ij\rangle\alpha\beta} 
J_{ij}^{\alpha\beta}\hat{\sigma}_{i}^{\alpha}\hat{\sigma}_{j}^{\beta}.$
Since there is no symmetry to constrain the interaction within each
bond, the NN inter-dimer interaction takes the form

\begin{align}
H_{J} & =\sum_{\langle ij\rangle}\left(\hat{\sigma}_{i}^{1},\hat{\sigma}_{i}^{2},\hat{\sigma}_{i}^{3}\right)\begin{pmatrix}J^{11} & J^{12} & \eta_{ij}J^{13}\\
J^{21} & J^{22} & \eta_{ij}J^{23}\\
\eta_{ij}J^{31} & \eta_{ij}J^{32} & J^{33}
\end{pmatrix}\begin{pmatrix}\hat{\sigma}_{j}^{1}\\
\hat{\sigma}_{j}^{2}\\
\hat{\sigma}_{j}^{3}
\end{pmatrix}\nonumber \\
 & \equiv\sum_{\langle ij\rangle}\hat{\boldsymbol{\sigma}}_{i} 
 \cdot\mathsf{J}_{ij}\cdot\hat{\boldsymbol{\sigma}}_{j}.\label{eq:inter-dimer}
\end{align}
Here $\hat{\boldsymbol{\sigma}}_{i}=\left(\hat{\sigma}_{i}^{1}, 
\hat{\sigma}_{i}^{2},\hat{\sigma}_{i}^{3}\right)^{T}$,
$\mathsf{J}_{ij}$ denotes the interaction matrices of the $\langle ij\rangle$
bond, $\langle ij\rangle$ follows the bond direction $i\rightarrow j$
as shown in Fig. \ref{fig:crystal}\textbf{c}, and $\eta_{ij}$ takes
the value $\pm1$ for the red and blue bonds as in Fig. \ref{fig:crystal}\textbf{c},
respectively. Note that for the NN inter-dimer interaction, the DM
interaction is allowed by symmetry. Moreover, there can be two types
of inter-dimer DM interactions: the bond-independent one $H_{DM}^{(1)}=\sum_{\langle ij\rangle}\mathbf{D}^{(1)}\cdot\left(\hat{\boldsymbol{\sigma}}_{i}\times\hat{\boldsymbol{\sigma}}_{j}\right)$
and the bond-dependent one $H_{DM}^{(2)}=\sum_{\langle ij\rangle}\mathbf{D}_{ij}^{(2)}\cdot\left(\hat{\boldsymbol{\sigma}}_{i}\times\hat{\boldsymbol{\sigma}}_{j}\right)$,
with $\mathbf{D}^{(1)}=(0,0,D^{z})$ and $\mathbf{D}_{ij}^{(2)}=\eta_{ij}(D^{x},D^{y},0)$.

%\subsect{Coupling to magnetic field. }
Finally we consider the Zeeman coupling of local moments to the external magnetic fields $H_\textrm{Zeeman}$. 
it is shown that all the three spin components are magnetic dipoles
that directly couple to the external magnetic field. As the dipole
axes of different sites only depend on their sublattices indices,
for simplicity here we only consider four rare-earth sites within
a unit cell, where each site is labeled by 
its sublattice index
$i$ ($i=0,1,2,3$). The general form of the Zeeman coupling takes
the form
\begin{equation}
H_{\textrm{Zeeman}}=-\sum_{i=0}^{3}\mu_{B}\mathbf{B}\cdot\mathsf{g}_{i}\cdot\hat{\boldsymbol{\sigma}_{i}},
\end{equation}
where $\left(\mathsf{g}_{i}\right)_{\alpha\beta}=g_{J}A^{(\beta)}n_{i,\alpha}^{(\beta)}$
is the $g$-tensor for the effective spins. Due to the site-dependence
of the local dipole axes $\mathbf{n}_{i}^{(\alpha)}$, the external
magnetic field couples to the effective spins in some complicated manner
as follows

\begin{align}
H_{[001]} & =-\mu_{B}g_{J}B^{[001]}\sum_{i=0}^{3} 
\left(A_{\perp}^{(1)}\hat{\sigma}_{i}^{1} 
+A_{\perp}^{(2)}\hat{\sigma}_{i}^{2}\right),\label{eq:001 zeeman}
\end{align}

\begin{align}
H_{[110]} & =-\mu_{B}g_{J}B^{[110]}\times\nonumber \\
 & \left[A_{\parallel}^{(1)}\left(\hat{\sigma}_{3}^{1} 
 -\hat{\sigma}_{1}^{1}\right)+A_{\parallel}^{(2)} 
 \left(\hat{\sigma}_{3}^{2}-\hat{\sigma}_{1}^{2}\right)  
 +A^{(3)}\left(\hat{\sigma}_{2}^{3}-\hat{\sigma}_{0}^{3}\right)\right], 
 \label{eq:110 zeeman}
\end{align}

\begin{align}
H_{[\bar{1}10]} & =-\mu_{B}g_{J}B^{[\bar{1}10]}\times\nonumber \\
 & \left[A_{\parallel}^{(1)}\left(\hat{\sigma}_{2}^{1} 
 -\hat{\sigma}_{0}^{1}\right)+A_{\parallel}^{(2)}\left(\hat{\sigma}_{2}^{2} 
 -\hat{\sigma}_{0}^{2}\right)+A^{(3)}\left(\hat{\sigma}_{1}^{3} 
 -\hat{\sigma}_{3}^{3}\right)\right],\label{eq:-110 zeeman}
\end{align}
where $A_{\parallel}^{(\alpha)}\equiv A^{(\alpha)}\cos\theta_{\alpha}$
and $A_{\perp}^{(\alpha)}\equiv A^{(\alpha)}\sin\theta_{\alpha}$
correspond to the magnitudes of the in-plane and out-of-plane components
of the dipole moments $\hat{\sigma}^{\alpha}$ ($\alpha=1,2$), respectively.
As shown above, the out-of-plane and in-plane magnetic fields couple
to the uniform and staggered dipolar magnetization, respectively.
The different ways of coupling will have profound implications 
for the behaviors of magnetization of the ``singlet'' and ``triplet''
dimer phases, as will be discussed later. Meanwhile, as the external
magnetic field always couples to multiple components of effective spins,
the high temperature magnetic susceptibilities would generally exhibit
non-Curie-Weiss behaviors. As such non-Curie-Weiss behaviors are difficult
to analyze in experiments, we will not discuss them in detail here
in our present work.

\subsect{Strong SOC dimer phases}\label{sec:Dimer-SOC}
To explore the effects of the strong SOC, we focus on a simple
limit, where intra-dimer interactions $H_{J^{\prime}}$
dominate over the inter-dimer interactions $H_{J}$. For the isotropic
(Heisenberg) SS model, the ground state at this limit is the dimer singlet 
phase, which is particularly notable for
its exact solvability and the presence of nearly localized triplon
excitations \cite{mcclarty2017topological}. However, the fate of %such 
these properties under strong SOC
remains unclear. To gain insight into the fate of the dimer phase %phases 
in the presence of strong SOC,  
we first %consider the strong intra-dimer limit where the inter-dimer 
%interactions are neglected, so that 
assume the ground state %becomes 
to be a product state of decoupled dimers. Note that the triplet state of a 
dimer splits  
%Due to 
in the presence of the XYZ anisotropy, so that the  
four eigenstates of a
%within each 
dimer are in general non-degenerate, and each can be the ground state.
%which are labeled as

Denoting these four states as
\begin{align}
|s\rangle & =\frac{1}{\sqrt{2}}\left(|\uparrow\downarrow\rangle-|\downarrow\uparrow\rangle\right),\\
|t_{1}\rangle & =\frac{-1}{\sqrt{2}}\left(|\uparrow\uparrow\rangle-|\downarrow\downarrow\rangle\right),\\
|t_{2}\rangle & =\frac{i}{\sqrt{2}}\left(|\uparrow\uparrow\rangle+|\downarrow\downarrow\rangle\right),\\
|t_{3}\rangle & =\frac{1}{\sqrt{2}}\left(|\uparrow\downarrow\rangle+|\downarrow\uparrow\rangle\right),
\end{align}
with energies $\epsilon_{s}=-\left(J^{\prime11}+J{}^{\prime22}+J{}^{\prime33}\right)/4$
and $\epsilon_{t_{\alpha}}=-\epsilon_{s}-J{}^{\prime\alpha\alpha}/2$
($\alpha=1,2,3$). Note that here $|t_{\alpha}\rangle$ is chosen
%as 
to be the eigenstate of the total spin operator 
$\hat{T}^{\alpha}\equiv\hat{\sigma}_{1}^{\alpha}+\hat{\sigma}_{2}^{\alpha}$:
$\hat{T}^{\alpha}|t_{\alpha}\rangle=0$. Following the convention
of isotropic Heisenberg dimers, here we still denote $|s\rangle$
as ``singlet'' dimer and $|t_{\alpha}\rangle$ as ``triplet''
dimers respectively, despite that all the energy levels are no longer
degenerate.

\begin{figure}
	\includegraphics[width=0.95\columnwidth]{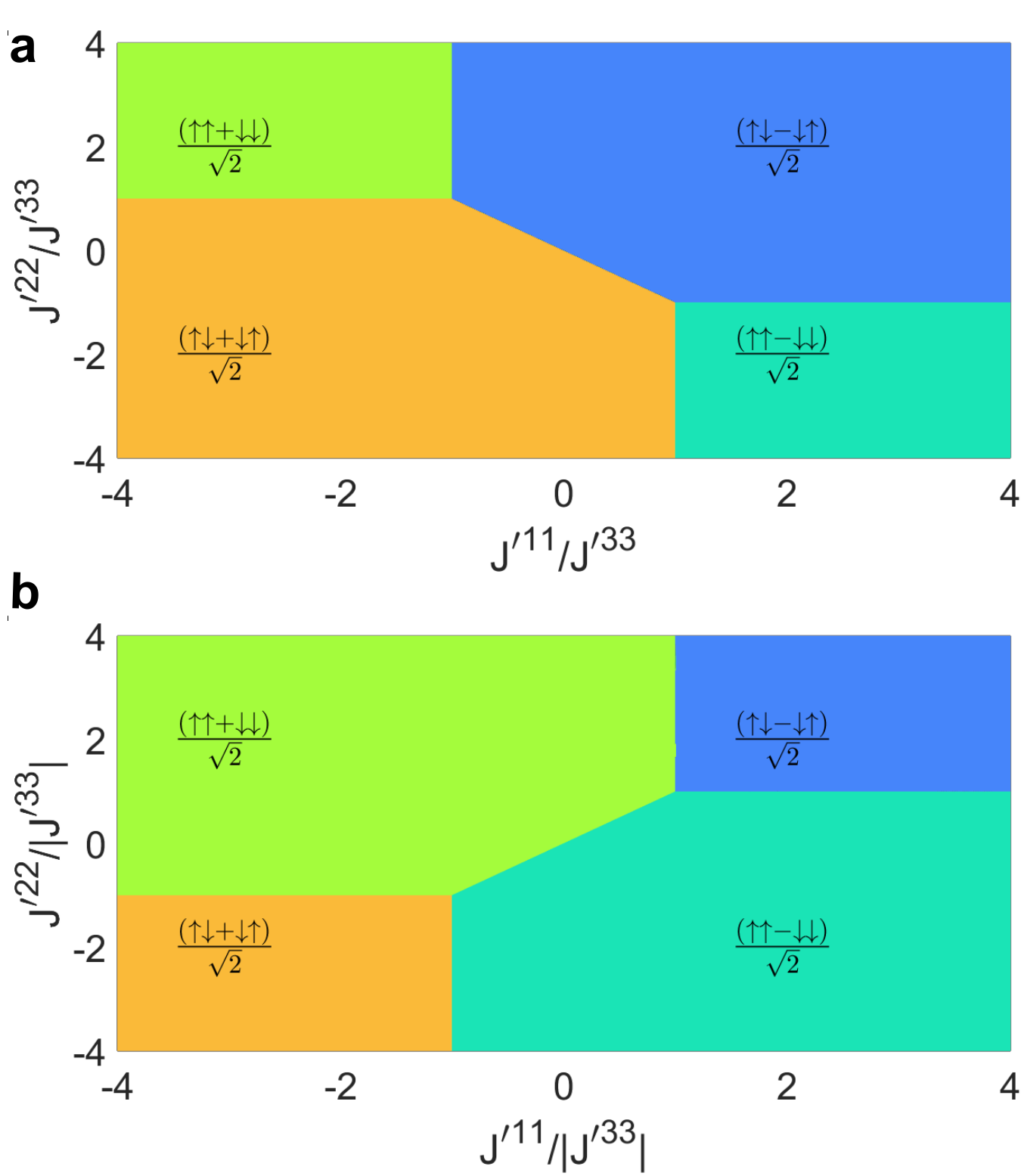} 
	\caption{ \textbf{Phase diagram of the effective model Eq. \eqref{eq:ham} in
			the intra-dimer limit $\mathsf{J}_{ij}=0$.} 
		%In panels \textbf{a} and \textbf{b}, $J^{\prime33}$ is set to be both antiferromagnetic and ferromagnetic, respectively.
		\textbf{a} antiferromagnetic $J^{\prime33}>0$ and \textbf{b} ferromagnetic $J^{\prime33}<0$. }
	\label{fig:phase_diagram}
\end{figure}

The phase diagram in the intra-dimer limit is shown in Fig. \ref{fig:phase_diagram}.
Depending on the values of $J^{\prime\alpha\alpha}$, %any
each of the four
dimer states can be the ground state of the system. Hence 
%for the whole system 
in principle we can have four types of dimerized ground states, 
$\ket{\psi_{s}}$
and $\ket{\psi_{t_{\alpha}}}$, as described by the product state
of $|s\rangle$ and $|t_{\alpha}\rangle$ on each dimer, respectively.
The fact that ``triplet'' dimer states are physically distinct from
the singlet ones can be derived from symmetries. In fact, for each
dimer, the singlet and triplet states transform differently under
spatial symmetries

\[
\sigma_{v}:|s\rangle\rightarrow+|s\rangle,|t_{1}\rangle\rightarrow-|t_{1}\rangle,|t_{2}\rangle\rightarrow-|t_{2}\rangle,|t_{3}\rangle\rightarrow+|t_{3}\rangle,
\]

\[
\sigma_{v}^{\prime}:|s\rangle\rightarrow-|s\rangle,|t_{1}\rangle\rightarrow-|t_{1}\rangle,|t_{2}\rangle\rightarrow-|t_{2}\rangle,|t_{3}\rangle\rightarrow+|t_{3}\rangle.
\]
The different symmetry representations of singlet and triplet states
forbids their mutual mixing as long as the crystalline symmetries
are preserved. Therefore, singlet and triplet dimer phases remain
symmetry distinct phases even when the inter-dimer interactions $H_{J}$
is introduced. Also note that $|t_{1}\rangle$ and $|t_{2}\rangle$
triplet states share the same symmetry representation and this is
because the effective-spin components $\hat{\sigma}^{1}$ and $\hat{\sigma}^{2}$
transform in the same way under spatial symmetries.

%\section{Experimental signatures of singlet and triplet dimer phases}

We have shown that two distinct dimerized
phases, the singlet and triplet dimer phases, can be stabilized in
the ground-state phase diagram and can be characterized by different 
representations
under spatial symmetries. The singlet dimer phase is usually stabilized 
when the intra-dimer interaction is antiferromagnetic and of either Heisenberg 
or XXZ-type.
It has been widely studied in systems with SS and/or bilayer lattices 
%\cite{mcclarty2017topological,fukumoto2000two,lohofer2015dynamical,zhang2024large}
, where 
physics such
as magnon Bose-Einstein condensation and the Higgs excitations are
involved \cite{mcclarty2017topological,fukumoto2000two,lohofer2015dynamical,zhang2024large,zapf2014bose,totsuka2001low}. In contrast, triplet dimer phases have not been investigated
in prior studies, to our knowledge. In the following, we discuss the
experimental signatures of singlet and triplet dimer phases. A comparison
on distinct experimental behaviors of singlet and triplet dimer phases
is summarized in Tab. \ref{tab:compare}.

\begin{table*}[t]
	\begin{ruledtabular}
		\begin{tabular}{ccc}
			& Singlet dimer phase & Triplet dimer phases\tabularnewline
			\colrule Zero-magnetization plateau & Present & Absent\tabularnewline
			Zero-field magnetic susceptibility & Suppressed & Pronounced\tabularnewline
			$M$ vs $B$ curve & Discontinuous & Continuous and smooth\tabularnewline
			$\Delta E$ vs $B$ curve & Non-analytical with gap closure & Smooth without gap closure\tabularnewline
			Excitation dispersion & Less dispersive & More dispersive\tabularnewline
		\end{tabular}
	\end{ruledtabular}
	
	\caption{\textbf{Comparison of experimental signatures for the singlet and
			triplet dimer phases at low temperatures.} Here we assume that the
		magnetic field is applied along to the {[}001{]} direction. Here we
		state that the dispersion within the triplet dimer phases are \textquotedblleft more
		dispersive\textquotedblright{} than singlet ones, as all the entries
		in $H_{J}$ contribute to dispersion above the triplet dimer phases
		at the linear order, while for the singlet dimerized phases only off-diagonal
		entries $J^{13}$, $J^{23}$, $J^{31}$ and $J^{32}$ contribute dispersion
		at the linear order.}
	\label{tab:compare}
\end{table*}

\subsect{Stability of dimer phases under small fields}
We first discuss the stability of the singlet and triplet dimer phases
under a small external magnetic field. To simplify our discussions,
we assume that the system lies deep 
in the dimer phase and the
inter-dimer interactions can be ignored compared to the intra-dimer
ones. As 
shown in Eqs.~\eqref{eq:001 zeeman}-\eqref{eq:-110 zeeman}, in our Kramers 
systems the coupling
to the magnetic field takes complicated forms, with the out-of-plane
and in-plane field components coupled to the uniform and staggered
dipolar magnetizations, respectively. In the following we will discuss
these two cases separately.

Note that within each unit cell there are two dimers denoted by the
dimer indices $\delta=A,B$. When an external magnetic field is applied
along the out-of-plane {[}001{]} direction, each dimer couples to the
external field in a uniform way according to Eq. (\ref{eq:001 zeeman}).
Therefore, it is sufficient to consider only 
a single dimer, say,
the $\delta=B$ dimer formed by RE$_{0}$ and RE$_{2}$. The Zeeman
coupling along the $[001]$ field direction takes the form
\begin{align}
H_{[001]} & 
=-\mu_{B}g_{J}B^{[001]}\sum_{i=0,2}\left(A_{\perp}^{(1)}\hat{\sigma}_{i}^{1} 
+A_{\perp}^{(2)}\hat{\sigma}_{i}^{2}\right),
\end{align}
and is represented in the matrix form with the dimer basis $\psi\equiv\begin{pmatrix}|s\rangle & |t_{1}\rangle & |t_{2}\rangle & |t_{3}\rangle\end{pmatrix}^{T}$:

\begin{align}
H_{[001]} & =-\mu_{B}g_{J}B^{[001]}\left(\begin{array}{cccc}
0\\
 & 0 &  & iA_{\perp}^{(2)}\\
 &  & 0 & -iA_{\perp}^{(1)}\\
 & -iA_{\perp}^{(2)} & iA_{\perp}^{(1)} & 0
\end{array}\right).
\end{align}
In the presence of the magnetic field $\parallel[001]$, the three
triplet states $|t_{\alpha}\rangle$ will mix with each other. Meanwhile,
the singlet state $|s\rangle$ cannot mix with triplet states as
is forbidden by symmetries: $H_{[001]}$ is odd under both $\sigma_{v}$
and $\sigma_{v}^{\prime}$; The mixing between $\ket{s}$ and $\ket{t_{3}}$
is forbidden by $\sigma_{v}^{\prime}$, while mixing between $\ket{s}$
and $\ket{t_{1}}$ ($\ket{t_{2}}$) is forbidden by $\sigma_{v}$.
Therefore, the singlet state $|s\rangle$ remains unaffected as an
eigenstate of the dimer system. If the singlet state is the zero-field
ground state, it will remain stable within a certain range of magnetic
fields, forming an approximate zero magnetization plateau with suppressed
magnetic susceptibility at low temperatures. On the other hand, if
the zero-field ground state belongs to the triplet dimer states $\ket{\psi_{t_{\alpha}}}$,
an infinitesimal field can already induce a non-zero magnetization.
Therefore, we expect a pronounced low-temperature magnetic susceptibility
for the triplet dimer states.

\begin{figure}
	\includegraphics[width=1\columnwidth]{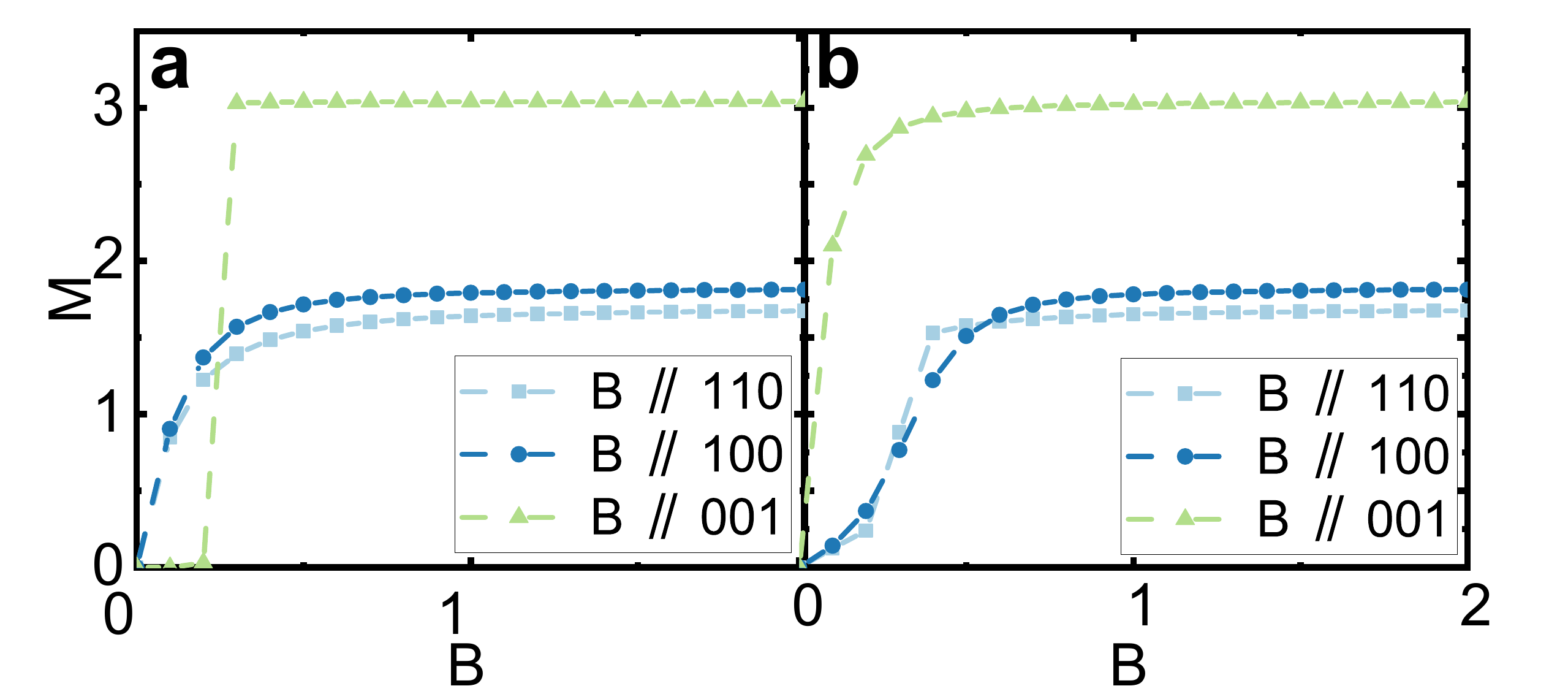}
	
	\caption{\textbf{Magnetization process of dimer phases upon external magnetic
			fields}. \textbf{a} singlet dimer phase $|s\rangle$, with parameters
		$J'^{11}=1.4$, $J'^{22}=1.7$, $J'^{33}=1$. \textbf{b} triplet dimer
		phase $|t_{3}\rangle$, with parameters $J'^{11}=-1.4$, $J'^{22}=-1.7$,
		$J'^{33}=1$. In both cases we set parameters $\mu_{B}g_{J}A_{\parallel}^{(1)}=1.1$,
		$\mu_{B}g_{J}A_{\parallel}^{(2)}=2.1$, $\mu_{B}g_{J}A_{\parallel}^{(3)}=1.0$,
		$\mu_{B}g_{J}A_{\perp}^{(1)}=2.2$, $\mu_{B}g_{J}A_{\perp}^{(2)}=2.1$.
		The temperature is set to $T=0.1J'^{33}$. The calculations are performed
		using the exact diagonalization of decoupled dimers.}
	\label{fig:MC}
\end{figure}

Then we consider the in-plane magnetic field, say, along the $[110]$
direction. In this case, the two dimers labeled by $\delta=A,B$ couple
to the magnetic field differently, so we will discuss them separately.
We first consider the $\delta=B$ dimer formed by RE$_{0}$ and RE$_{2}$.
The Zeeman coupling is

\begin{equation}
H_{[110]}^{(B)}=-\mu_{B}g_{J}B^{[110]}A_{\parallel}^{(3)}\left(\hat{\sigma}_{2}^{3}-\hat{\sigma}_{0}^{3}\right),
\end{equation}
with the matrix representation in the $\psi$ basis

\begin{equation}
H_{[110]}^{(B)}=\mu_{B}g_{J}B^{[110]}\left(\begin{array}{cccc}
0 &  &  & A_{\parallel}^{(3)}\\
 & 0\\
 &  & 0\\
A_{\parallel}^{(3)} &  &  & 0
\end{array}\right).
\end{equation}
In this case, the singlet $|s\rangle$ will mix with $|t_{3}\rangle$,
while $|t_{1}\rangle$ and $|t_{2}\rangle$ remain unaffected by the
field. Meanwhile, for the $\delta=A$ dimer formed by RE$_{1}$ and
RE$_{3}$, the matrix representation of the Zeeman coupling takes
the form

\begin{equation}
H_{[110]}^{(A)}=-\mu_{B}g_{J}B^{[110]}\left[A_{\parallel}^{(1)}\left(\hat{\sigma}_{3}^{1}-\hat{\sigma}_{1}^{1}\right)+A_{\parallel}^{(2)}\left(\hat{\sigma}_{3}^{2}-\hat{\sigma}_{1}^{2}\right)\right],
\end{equation}
with the matrix representation
\begin{equation}
H_{[110]}^{(A)}=\mu_{B}g_{J}B^{[110]}\left(\begin{array}{cccc}
0 & A_{\parallel}^{(1)} & A_{\parallel}^{(2)} & 0\\
A_{\parallel}^{(1)} & 0\\
A_{\parallel}^{(2)} &  & 0\\
0 &  &  & 0
\end{array}\right).
\end{equation}
Therefore, for the $\delta=A$ dimer, the singlet $|s\rangle$ will
mix with $|t_{1}\rangle$ and $|t_{2}\rangle$, while $|t_{3}\rangle$
remains unaffected. Combining the magnetization process of the two
dimers, we conclude that under an in-plane magnetic field, all the
dimer states will become unstable and become magnetized by the field,
hence the whole system will not exhibit any zero magnetization plateau
(Fig. \ref{fig:MC}).

\begin{figure}
	\includegraphics[width=1\columnwidth]{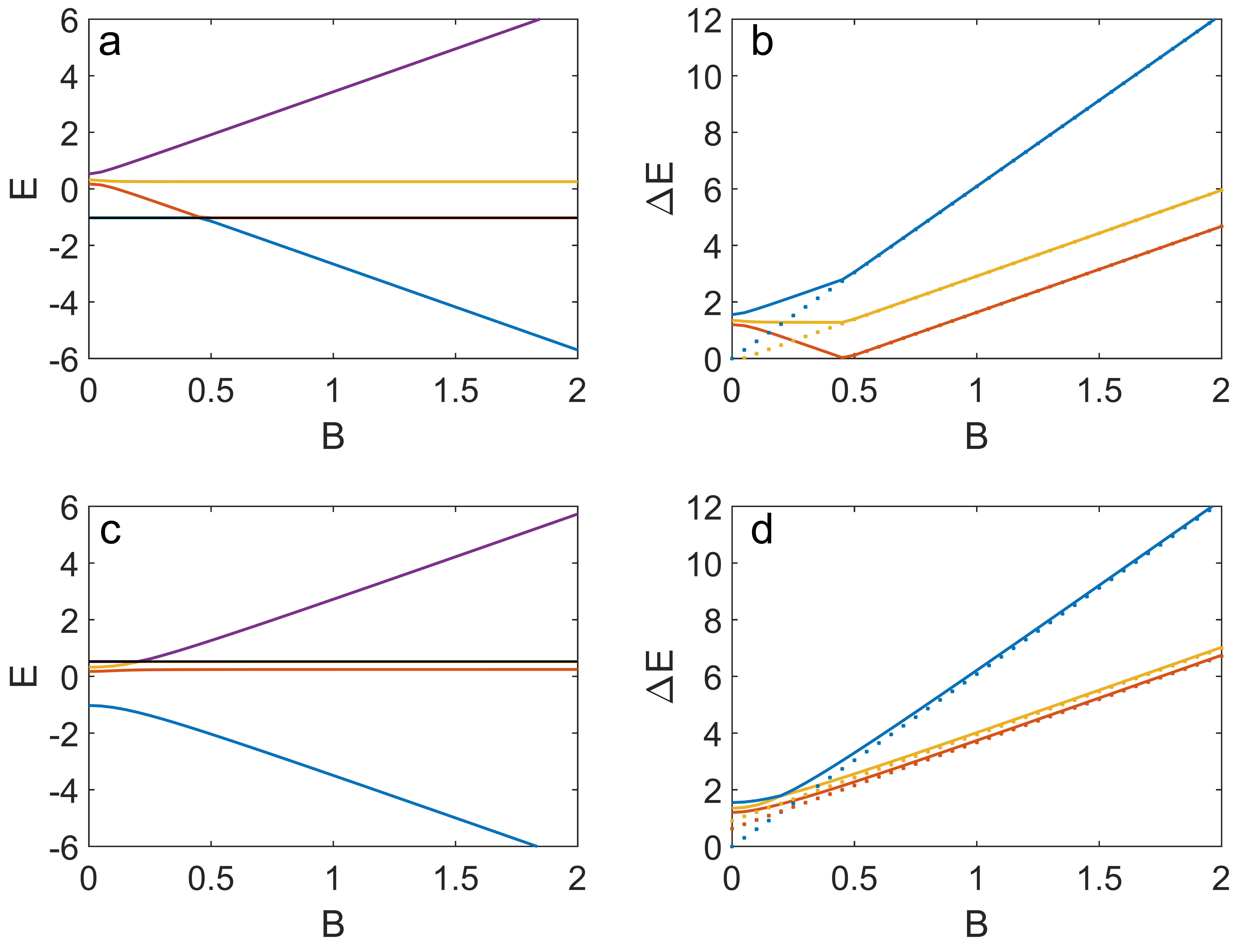} 
	\caption{\textbf{ Evolution of energy spectra under external magnetic field
			$\parallel[001]$. }Energy levels and transition energies within \textbf{a,b}
		the singlet dimer phase, and \textbf{c,d} the triplet dimer phase.
		In \textbf{a,c} the the singlet energy level is marked by the bold
		black line, while the three triplet levels are marked by colored solid
		lines. In \textbf{b,d} the high-field asymptotic behaviors of three
		excitations are indicated by dotted lines. The system parameters are
		identical to that in Fig. \ref{fig:MC}, except that the temperature
		$T$ is set to zero.}
	
	\label{fig:EvsB}
\end{figure}

\subsect{Spectral evolution under magnetic field $\parallel[001]$} 
Here we discuss the evolution of excitation spectra under external
magnetic fields that can be investigated through neutron scattering,
terahertz spectroscopy and electron spin resonance experiments. To
simplify our discussions, we consider that the magnetic field is applied
along the $[001]$ direction, \textit{i.e.}, $\mathbf{B}=B\hat{z}$. We
also assume that the system lies deep within the dimer phase and the
inter-dimer interactions can be ignored compared to the intra-dimer
ones. As previously mentioned, under magnetic fields $\parallel[001]$,
the mixing between the singlet and triplet levels is forbidden by
symmetries. The energy of the singlet state $\ket{s}$ remains unaffected
in the presence of the magnetic field:
\begin{equation}
E_{s}(B)=\epsilon_{s}.
\end{equation}
In contrast, the three triplet states experience mutually mixing in
presence of magnetic fields. Such mutual mixing will result avoided
level crossing of triplets under magnetic field. However, level crossing
between the singlet and triplet levels still are allowed by symmetries.
We denote the three triplet eigenstates as $\ket{t,-}$, $\ket{t,0}$,
$\ket{t,+}$ with energies $E_{t,-}(B)<E_{t,0}(B)<E_{t,+}(B)$. In
the limit where the field dominates over the exchange interactions,
the energies of the three triplet levels $\ket{t,\pm}$, $\ket{t,0}$
can be obtained from the non-degenerate perturbation theory:

\begin{align}
E_{t,\pm}(B) & =\pm\mu_{B}g_{J}B\sqrt{\left(A_{\perp}^{(1)}\right)^{2}+\left(A_{\perp}^{(2)}\right)^{2}}+E_{t,\pm}^{(1)}+O(1/B),\\
E_{t,0}(B) & =E_{t,0}^{(1)}+O(1/B),
\end{align}
where
\begin{align}
E_{t,\pm}^{(1)} & =\frac{1}{2}\left[\frac{\left(A_{\perp}^{(2)}\right)^{2}\epsilon_{t_{1}}+\left(A_{\perp}^{(1)}\right)^{2}\epsilon_{t_{2}}}{\left(A_{\perp}^{(1)}\right)^{2}+\left(A_{\perp}^{(2)}\right)^{2}}+\epsilon_{t_{3}}\right],\\
E_{t,0}^{(1)} & =\frac{\left(A_{\perp}^{(1)}\right)^{2}\epsilon_{t_{1}}+\left(A_{\perp}^{(2)}\right)^{2}\epsilon_{t_{2}}}{\left(A_{\perp}^{(1)}\right)^{2}+\left(A_{\perp}^{(2)}\right)^{2}}
\end{align}
are the first-order perturbations that exhibit no field dependence,
and $O(1/B)$ comes from the second-order corrections from the perturbation
theory. Therefore, in the high field limit, the three branches of
excitations involving transitions from the lowest $\ket{t,-}$ state
to the excited $\ket{t,+}$, $\ket{t,0}$ and $\ket{s}$ levels should
satisfy the following linear asymptotic behaviors:

\begin{align}
\Delta E_{1}(B) & =2\mu_{B}g_{J}B\sqrt{\left(A_{\perp}^{(1)}\right)^{2}+\left(A_{\perp}^{(2)}\right)^{2}}+O(1/B),\\
\Delta E_{2}(B) & =\mu_{B}g_{J}B\sqrt{\left(A_{\perp}^{(1)}\right)^{2}+\left(A_{\perp}^{(2)}\right)^{2}}+\left(E_{t,0}^{(1)}-E_{t,\pm}^{(1)}\right)+O(1/B),\\
\Delta E_{3}(B) & =\mu_{B}g_{J}B\sqrt{\left(A_{\perp}^{(1)}\right)^{2}+\left(A_{\perp}^{(2)}\right)^{2}}+\left(\epsilon_{s}-E_{t,\pm}^{(1)}\right)+O(1/B).
\end{align}
Due to the absence of symmetry-based selection rules, all the three
branches are generally visible in the neutron scattering, terahertz
spectroscopy or electron spin resonance experiments. By extrapolating
the high-field behavior of the three excitations, we find that the
branch 1 has twice the slope compared to the branch 2 and 3 with vanishing
intercept along the $B=0$ axis. Meanwhile, branch 2 and 3 identical
slopes with non-vanishing intercept along the $B=0$ axis. From the
evolution of energy spectrum under external magnetic fields, the experimental
signatures of the singlet and triplet dimer phases are easily distinguishable:
\begin{itemize}
\item For the singlet dimer phase, the singlet level remains as the ground
state at small fields, hence there must be a level crossing from singlet
$\ket{s}$ to triplet $\ket{t,-}$ at a finite critical field $B_{c}$.
This level crossing is expected to manifest as a non-analytical behavior
of the excitation branch 3, where the gap closes and reopens across
$B_{c}$, see Fig. \ref{fig:EvsB}\textbf{a,b}. A jump of magnetization
across $B_{c}$ is also expected;
\item For the triplet dimer phase, the zero-field ground state is adiabatically
connected to the high-field phase, meaning that a phase transition
may not necessarily present. As a result, the gap-closing and reopening
behavior observed in the singlet dimer phase is not expected. Instead,
the excitation energies evolve smoothly with magnetic fields, exhibiting
the high-field behaviors as discussed previously. See Fig. \ref{fig:EvsB}\textbf{c,d}.
\end{itemize}

\vspace{5cm}

\subsect{Effects of inter-dimer interactions}
Our previous discussions on the dimer phases are restricted to the
intra-dimer limit, where the inter-dimer interactions are neglected
and the ground states are described by product states of decoupled
dimers. The inter-dimer interactions $H_{J}$ introduces correlation between dimers,
which can give rise to various field-induced intermediate 
phases as has been observed in both the isotropic 
SrCu$_2$(BO$_3$)$_2$ system and the anisotropic XXZ/XYZ models~\cite{nonkramers}. 
The aspect of field-induced intermediate phases will be further discussed in \textit{Discussion}. On the other hand, the inter-dimer correlation can
lead to correlated quantum dynamics involving nearby dimers, which 
creates quantum entanglement between dimers and generate dispersions 
to the elementary excitations, and will be discussed as follows. 

The quantum dynamics above the dimerized ground state
have been extensively studied in
the context of the isotropic (Heisenberg) SS model: It was found that
the inter-dimer interactions $H_{J}$ have vanishing effects to the
dimer ground state, resulting in the emergent exact solvability of the
dimerized phase: the dimer singlet product state $\ket{\psi_{s}}$
remains to be the exact ground state of the system when 
$J/J^{\prime}\lesssim0.675$
\cite{lee2019signatures,xi2023plaquette}. Since the quantum entanglement
of the dimer ground state is restricted within each dimer, the low-lying
quantum excitations must be highly localized, resulting in a nearly flat 
triplet band in the spin excitation spectrum with %Indeed, the single triplet 
%excitations are nearly flat, with 
	very weak dispersion caused by the
sixth and higher order terms in $J/J^{\prime}$ 
\cite{miyahara1999exact,miyahara2000superstructures,kageyama2000direct}.

Then we move %away from the isotropic limit and 
on to incorporate the strong
SOC. We adopt the bond-operator formalism \cite{lauchli2002phase,mcclarty2018topological,sachdev1990bond}
to analyze the quantum dynamics of the dimerized phases. In this formalism,
each dimer state $|s\rangle$ or $|t_{\alpha}\rangle$ can be expressed
by creating a bond-operator boson out of the vacuum $|\textrm{vac}\rangle$,
\textit{i.e.},
\[
|s\rangle=\hat{s}^{\dagger}|\textrm{vac}\rangle,\,|t_{\alpha}\rangle=\hat{t}_{\alpha}^{\dagger}|\textrm{vac}\rangle.
\]
Then, each spin operator can be expressed as quadratics of bond operators,
see Methods for details. Note that the above bond-operator descriptions
are limited to a single dimer. In the SS lattice, each unit cell contains
two dimers, so each dimer can be uniquely labeled by its unit cell
position $\mathbf{r}$ and the dimer index $\delta=A,B$, as shown
in Fig. \ref{fig:crystal}\textbf{c}. In the following, we discuss
the quantum dynamical behaviors of the singlet and triplet dimer phases
separately. We will show that only off-diagonal entries $J^{13}$,
$J^{23}$, $J^{31}$ and $J^{32}$ in $H_{J}$ contribute dispersion
for the singlet dimerized phase, while all the entries in $H_{J}$
contribute to dispersion above the triplet dimer phases.

\hide{As has been discussed previously, the ``singlet'' dimer state,
corresponding to the product state of singlet $|s\rangle$ within
each dimer $|\psi_{s}\rangle\equiv\prod_{\mathbf{r}}\hat{s}_{\mathbf{r},A}^{\dagger}\hat{s}_{\mathbf{r},B}^{\dagger}|\textrm{vac}\rangle$,
has been proved to be the exact eigenstate of the isotropic Heisenberg
SS model in all parameter regime, and becomes the exact ground state
when $J/J^{\prime}\lesssim0.675$. For the anisotropic XYZ model we
considered here, the ``singlet'' state $|s\rangle$ remains to be
the ground state provided $J^{\prime\alpha\alpha}+J^{\prime\beta\beta}>0$
for all $\alpha,\beta=1,2,3$. Therefore, the ``singlet'' dimer
state can be stable even in the presence of significant XYZ anisotropy.}

\begin{figure}
	\includegraphics[width=1\columnwidth]{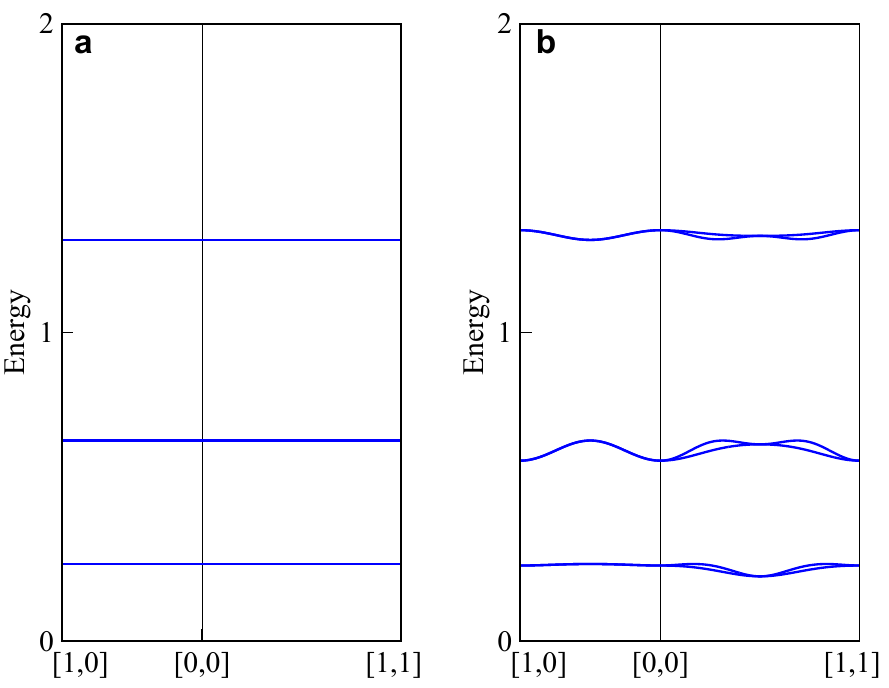}
	
	\caption{\textbf{Dispersion of excitations in the singlet dimer phase obtained
			by linear bond-operator theory}. The system parameters taken are $J^{\prime11}=1.7$,
		$J^{\prime22}=0.9$, $J^{\prime33}=-0.4$, $J^{11}=J^{22}=J^{33}=0.3$,
		$J^{12}=0.1$, $J^{21}=0$, \textbf{a} $J^{13}=J^{23}=J^{31}=J^{32}=0$
		that results a flat dispersion; \textbf{b} $J^{13}=0.1$, $J^{23}=0.13$,
		$J^{31}=0.05$, $J^{32}=-0.05$ that results dispersive excitations.}
	
	\label{fig:singlet_excitation}
\end{figure}

For the singlet dimer phase, we first consider a simple limit $J^{13}=J^{23}=J^{31}=J^{32}=0$,
making the interactions of the red and blue bond identical, \textit{i.e.},
$\mathsf{J}_{ij}=\mathsf{J}$. Then, the NN inter-dimer interactions
can be recast as interactions of each spin in one dimer with the total
spin of its adjacent dimer $\hat{\mathbf{T}}$,

\begin{align}
H_{J}=\sum_{\mathbf{r}}\,\, & \big[\hat{\boldsymbol{\sigma}}_{\mathbf{r},B,1}\cdot\mathsf{J}\cdot\hat{\mathbf{T}}_{\mathbf{r},A}+\hat{\boldsymbol{\sigma}}_{\mathbf{r},A,2}\cdot\mathsf{J}\cdot\hat{\mathbf{T}}_{\mathbf{r}+\mathbf{a},B}\nonumber \\
+ & \hat{\boldsymbol{\sigma}}_{\mathbf{r}+\mathbf{a},B,2}\cdot\mathsf{J}\cdot\hat{\mathbf{T}}_{\mathbf{r}+\mathbf{b},A}+\boldsymbol{\sigma}_{\mathbf{r}+\mathbf{b},A,1}\cdot\mathsf{J}\cdot\hat{\mathbf{T}}_{\mathbf{r},B}\big].
\end{align}
Note that the total spin operator $\hat{\mathbf{T}}$ vanishes when
acting upon the singlet dimer state: $\hat{\mathbf{T}}_{\mathbf{r}}\ket{\psi_{s}}=0$.
Therefore, we have

\begin{equation}
H_{J}\ket{\psi_{s}}=0,
\end{equation}
the singlet dimer state $|\psi_{s}\rangle$ remains to be an exact
eigenstate even in the presence of XYZ anisotropy. Meanwhile, the
excitations above the singlet dimerized state can be obtained by condensing
the singlet boson where a single boson $\hat{s}$ gains a non-vanishing
expectation value
\begin{equation}
\hat{s}=\hat{s}^{\dagger}\approx\sqrt{1-\sum_{\alpha}\hat{t}_{\alpha}^{\dagger}\hat{t}_{\alpha}},
\end{equation}
then the triplet can be regarded as elementary quantum excitations
in this dimerized phase. By expanding $H_{J}$ in terms of triplet
bosons order-by-order, we find that at the quadratic level no triplet
bilinears present, indicating that triplet excitations are almost
localized, see Fig. \ref{fig:singlet_excitation}\textbf{a}. By applying
perturbation theory, we find that the dispersion of triplet excitations
can only be caused by the sixth and higher orders of triplet operators
in $H_{J}$, analogous to the isotropic Heisenberg case \cite{miyahara2003theory,knetter2000dispersion,knetter2000symmetries}.

Away from the limit $J^{13}=J^{23}=J^{31}=J^{32}=0$, $H_{J}$ involves
second-order of the singlet operators $\sim t^{\dagger}st^{\dagger}s$,
as correlated tunneling of two nearby singlet dimers to triplet dimers.
On one hand, such correlated tunneling process makes the singlet dimer
product state $\ket{\psi_{s}}$ no longer the exact eigenstate of
the system. On the other hand, condensation of singlet bosons results
in direct hopping of triplet bosons at the first order of $H_{J}$.
Therefore, a non-zero $J^{13},J^{23},J^{31}$ or $J^{32}$ would contribute
to a more considerable bandwidth of single triplet excitations linear
in $H_{J}$, see Fig. \ref{fig:singlet_excitation}\textbf{b}.

\hide{If the ``triplet'' $|t_{\alpha}\rangle$ state has the lowest
energy within a dimer, the ``triplet'' dimer state, as described
by the variational wave functions of $|t_{\alpha}\rangle$ triplet
product state $|\psi_{t_{\alpha}}\rangle=\prod_{\mathbf{r}}\hat{t}_{\mathbf{r},A,\alpha}^{\dagger}\hat{t}_{\mathbf{r},B,\alpha}^{\dagger}|\textrm{vac}\rangle$,
will be stabilized as the ground state of the system in the $H_{J}=0$
limit. Stabilizing such triplet dimer state requires strong spin anisotropy,
\textit{i.e.}, it requires that at least one $J^{\prime\alpha\alpha}$
is ferromagnetic and has a larger absolute value than one of the antiferromagnetic
$J^{\prime\alpha\alpha}$ values. Similarly, the quantum dynamics
of the triplet dimer phase can be investigated by expressing $H_{J}$
in terms of the bond operators. }

\begin{figure}
	\includegraphics[width=1\columnwidth]{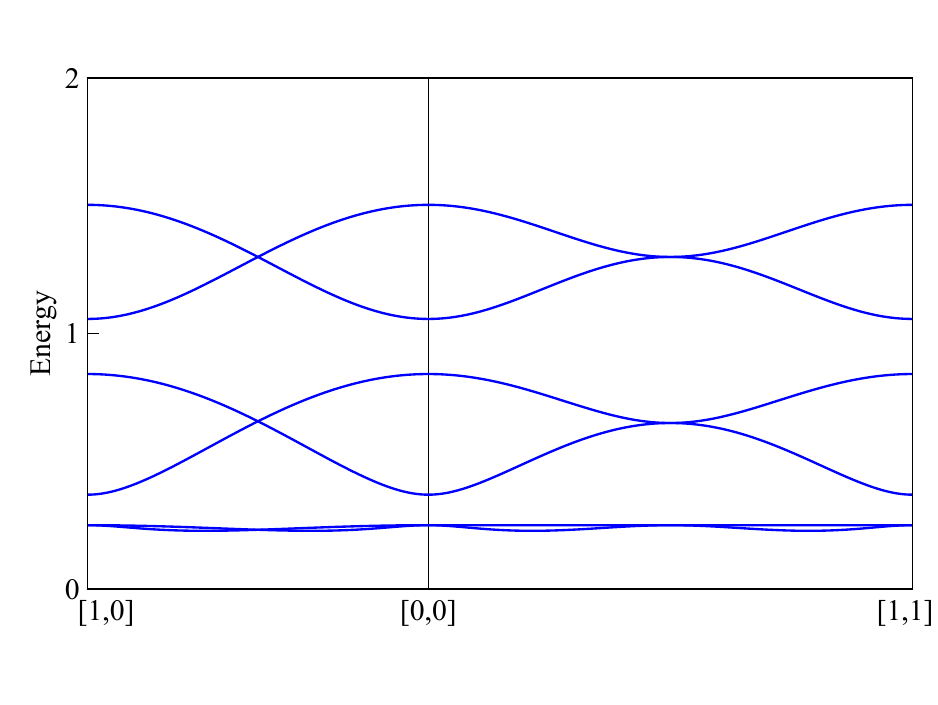}
	
	\caption{\textbf{Dispersion of excitations in the triplet dimer phase obtained
			from the linear bond-operator theory}. The system lies within the
		$\ket{\psi_{t_{2}}}$ phase with exchange parameters $J^{\prime11}=1.7$,
		$J^{\prime22}=-0.9$, $J^{\prime33}=0.4$, $J^{11}=J^{22}=J^{33}=0.11,J^{12}=0.07,J^{13}=J^{21}=J^{23}=J^{31}=J^{32}=0.$}
	
	\label{fig:triplet_excitation}
\end{figure}

Then we consider the quantum dynamics above the triplet dimer phases.
In the bond-operator representation, we find that all entries in $H_{J}$
contribute to terms of $t_{\beta}^{\dagger}t_{\beta}^{\dagger}t_{\alpha}t_{\alpha}$
($\beta\neq\alpha$) that drives the system away from the pure product
state $|\psi_{t_{\alpha}}\rangle$. Hence $|\psi_{t_{\alpha}}\rangle$
is not an exact eigenstate of the system if a finite inter-dimer coupling
$H_{J}$ is present. Meanwhile, the quantum excitations above the
triplet dimer state $\ket{\psi_{t_{\alpha}}}$ can be obtained by condensing $\hat{t}_{\alpha}$

\begin{equation}
\hat{t}_{\alpha}=\hat{t}_{\alpha}^{\dagger}\approx\sqrt{1-\hat{s}^{\dagger}\hat{s}-\sum_{\beta\neq\alpha}\hat{t}_{\beta}^{\dagger}\hat{t}_{\beta}}.
\end{equation}
The dispersion of quantum excitation is presented in Fig. \ref{fig:triplet_excitation}.
We note that in the triplet dimer phase, $\hat{t}_{\beta}$ ($\beta\neq\alpha$)
excitations are generally dispersive, in sharp contrast to the singlet
dimer phase. This is also because of the presence of $t_{\beta}^{\dagger}t_{\beta}^{\dagger}t_{\alpha}t_{\alpha}$
terms ($\beta\neq\alpha$) that appear in $H_{J}$.

\subsect{Implications for Yb$_{2}$Be$_{2}$GeO$_{7}$}
Here we discuss the implication of our results for the Yb$_{2}$Be$_{2}$GeO$_{7}$
compound \cite{YBGO,pula2024candidate}. In this system, thermodynamic
measurements reveal that magnetic entropy saturates to about $R\ln2$
at around 3 K, indicating that each Yb$^{3+}$ effectively behaves
a spin-1/2 system with only the lowest Kramers doublet being relevant
at low temperatures. Interestingly, no evidence of long-range magnetic
order has been observed in either specific heat or susceptibility
measurements down to 50 mK. Moreover, muon spin relaxation experiments
reveal persistent spin dynamics down to 17 mK \cite{pula2024candidate}.
These experimental observations seem to suggest a quantum spin liquid
ground state. However, the specific heat exhibits an activated behavior
at sufficiently low temperatures, implying the presence of a finite
gap  in this system. This gap is inconsistent with the characteristics
of most spin liquid candidate materials but more indicative of a dimerized
ground state. Furthermore, the nature of this dimerized phase can
be identified with the published data: The magnetization curve along
the $[001]$ field direction at $0.4$ K reveals the absence of a
phase transition or zero-magnetization plateau, with quite pronounced
magnetic susceptibility at weak fields \cite{YBGO,pula2024candidate}.
Such behavior is rather inconsistent with a singlet dimer state but
strongly suggests a triplet dimer ground state. This inference is
further supported by the electron spin resonance (ESR) measurements,
which show a smooth evolution of spectrum in all three excitation
modes, indicating the absence of gap closure in the presence of magnetic
field along $\parallel[001]$. In conclusion, all these comprehensive
findings collectively suggest that Yb$_{2}$Be$_{2}$GeO$_{7}$ is
more appropriately characterized as exhibiting a triplet dimer ground
state.

\sect{Discussion}%\label{sec:Discussions}

In this work, we develop a theory describing the quantum magnetism
of SS magnets with Kramers rare-earth ions. In contrast to the non-Kramers
systems where the local moments comprise both magnetic dipoles and
quadrupoles, for Kramers systems all the effective-spin components
are magnetic dipoles, with the dipole axes sublattice dependent. The
interactions between local moments are described by an extended XYZ
model on both intra- and inter-dimer bonds. We further illustrate
the distinct  thermodynamical and spectral signatures of singlet and
triplet dimer phases that can be experimentally probed,  as is summarized
in Tab. \ref{tab:compare}. We also suggest that the recently discovered
SS magnet Yb$_{2}$Be$_{2}$GeO$_{7}$ may host a triplet dimer ground-state.

This work on Kramers systems, together with the parallel study \cite{nonkramers}
on the non-Kramers counterpart, provides a comprehensive framework
of the underlying magnetism in the 2217 SS rare-earth compounds. Despite
the seeming similarity between Kramers and non-Kramers magnets, we
would like to emphasize that these two systems are fundamentally
different both in terms of symmetry and the resulting physical phenomena.
While the essential properties of non-Kramers magnets were hidden
quadrupole local moments and the intrinsic field, all those %such 
properties are absent in Kramers systems due to the different symmetry. 
On the other hand, the dimerized phase is a theme unique to Kramers 
systems. Even if they could appear in non-Kramers magnets,
they would be easily destabilized by the intrinsic field. 

Here we make a comparison of our strong SOC Kramers systems with the well-known SS magnet SrCu$_2$(BO$_3$)$_2$.
The the rare-earth compounds \textit{RE}$_2$Be$_2$\textit{X}O$_7$ (\textit{X}=Si, Ge) and the low-temperature phase of SrCu$_2$(BO$_3$)$_2$ share identical symmetries at the dimer center: both systems lack spatial inversion symmetry but retain two mirror symmetries parallel and perpendicular to the dimer direction. 
For pure spin systems such as SrCu$_2$(BO$_3$)$_2$ and Gd$_2$Be$_2$\textit{X}O$_7$, this symmetry implies an intra-dimer DM interaction in the global spin frame, with the DM vector $\mathbf{D}$ lying in-plane and perpendicular to the dimer. 
However, in most rare-earth SS magnets (\textit{RE}$\neq$Gd), the effective spins arise from strong SOC and are subjected to low-symmetry CEF environment, making spin components in the global coordinates no longer satisfy the canonical commutation relation. 
This complexity necessitates the adoption of local coordinate frames, where the moment directions of the canonical spin components become typically non-orthogonal and sublattice dependent. 
With the local-axes described in Fig.~\ref{fig:crystal}, the antisymmetric 
intra-dimer DM interaction transforms into a symmetric off-diagonal $\Gamma$ 
term, and can be eventually absorbed into the diagonal terms of the XYZ model via a global 
effective-spin rotation (see \textit{Effective Hamiltonian}). 
Consequently, no intra-dimer DM interaction remains within 
our formulation for effective spins under strong SOC.

We also note that triplet dimer materials Yb$_2$Be$_2$\textit{X}O$_7$ show no intermediate phases upon external magnetic field, in contrast to SrCu$_2$(BO$_3$) that host a series of field-induced phases, including a number of fractional magnetization plateaus. 
SrCu$_2$(BO$_3$)$_2$ belongs to the ``singlet'' dimer phase within our classificaton, and is very robust against inter-dimer interactions. 
The inter-dimer interaction of SrCu$_2$(BO$_3$)$_2$ is relatively large ($J\sim 0.63J^\prime$), in which the large inter-dimer correlation yields a cascade of platateaux transitions upon external field. 
In the strong SOC limit, the magnetization of such ``singlet'' dimer phase has been analyzed in our previous paper~\cite{nonkramers}, and many intermediate field-induced phases were also observed (note that in the context of non-Kramers systems, the field has an intrinsic origin rather than being externally applied). 
Meanwhile, our present study mainly focuses on the ``triplet'' dimer phase that are generally fragile against inter-dimer interactions. 
The relevant materials Yb$_2$Be$_2$\textit{X}O$_7$ also show vanishing inter-dimer interactions, which are too weak to produce any intermediate field-induced phases. 
Nevertheless, if the inter-dimer interactions increase and the system transitions out of the triplet dimer phase, various intermediate field-induced phases could still emerge.
For example, in Er$_2$Be$_2$GeO$_7$ the dipole moments are sizable and the long-range dipole-dipole interactions are more prominent. This inter-dimer correlations could help stabilize several plateau phases upon external magnetic fields, as observed in the recent experiment~\cite{EBGO}. 
However, this aspect goes beyond the scope of our current work. 

Finally, we emphasize that the concept of ``triplet''
dimers here in strong SOC SS magnets can be extended into a broader
context within quantum magnetism. First, the triplet dimerized phases
are not confined to the SS lattice and can be generalized to other
lattice geometries that support dimerized phases, such as honeycomb,
bilayer triangular, and bilayer square lattices \cite{nishimoto2016strongly,yu2024physics,moessner2001phase,laflorencie2009theory}. These systems with
various symmetries can provide fertile ground for exploring triplet
dimer physics in diverse settings. Second, this concept can be further
expanded to systems beyond dimerization to include scenarios where
spins group into larger clusters than dimers, such as four-spin groups
that organize into a ``triplet plaquette singlet''. While this broader
scope demonstrates the rich potential of triplet dimer physics, it
lies beyond the focus of our current work and is left for future investigation.

\sect{Methods}

\subsect{Bond-operator theory}
We study the dispersion of excitations in the dimer phase using the
bond-operator method \cite{lauchli2002phase,mcclarty2018topological,sachdev1990bond},
a powerful tool that investigates the quantum dynamics of dimerized
systems. In this formalism, each dimer state $|s\rangle$ or $|t_{\alpha}\rangle$
($\alpha=1,2,3$) can be expressed as applying bond-operator bosons
$\hat{s}^{\dagger}$ and $\hat{t}_{\alpha}^{\dagger}$ out of the
vacuum $|\textrm{vac}\rangle$, \textit{i.e.},
\[
|s\rangle=\hat{s}^{\dagger}|\textrm{vac}\rangle,\,|t_{\alpha}\rangle=\hat{t}_{\alpha}^{\dagger}|\textrm{vac}\rangle,
\]
The Hilbert space of the bosons is larger than the original dimer
Hilbert space and includes unphysical states. To limit the boson Hilbert
space to its physical sector, a hard-core constraint must be imposed
on each dimer,

\begin{equation}
\hat{s}^{\dagger}\hat{s}+\sum_{\alpha}\hat{t}_{\alpha}^{\dagger}\hat{t}_{\alpha}=1.
\end{equation}
The two spins within a dimer can be expressed as quadratics of bond
operators,

\begin{align}
\hat{\sigma}_{1}^{\alpha} & =+\frac{1}{2}\left(\hat{s}^{\dagger}\hat{t}_{\alpha}+\hat{t}_{\alpha}^{\dagger}\hat{s}\right)-\frac{i}{2}\epsilon_{\alpha\beta\gamma}\hat{t}_{\beta}^{\dagger}\hat{t}_{\gamma},\\
\hat{\sigma}_{2}^{\alpha} & =-\frac{1}{2}\left(\hat{s}^{\dagger}\hat{t}_{\alpha}+\hat{t}_{\alpha}^{\dagger}\hat{s}\right)-\frac{i}{2}\epsilon_{\alpha\beta\gamma}\hat{t}_{\beta}^{\dagger}\hat{t}_{\gamma},
\end{align}
Therefore we can rewrite the total Hamiltonian Eq. (\ref{eq:ham})
in terms of bond-operators.

Note that the total spin $\hat{T}^{\alpha}$ for each dimer can be
expressed as

\begin{equation}
\hat{T}^{\alpha}\equiv\hat{\sigma}_{1}^{\alpha}+\hat{\sigma}_{2}^{\alpha}=-i\epsilon_{\alpha\beta\gamma}\hat{t}_{\beta}^{\dagger}\hat{t}_{\gamma},
\end{equation}
that only involves hopping between triplets but not singlet. Meanwhile,
a single spin operator $\hat{\sigma}_{1,2}^{\alpha}$ involves both
triplet hopping process, as well as mixing between singlet and triplets.

The intra-dimer interaction $H_{J^{\prime}}$ is diagonal in the bond-operator
representation that can be recast as the chemical potential of bosons

\begin{equation}
H_{J^{\prime}}=\sum_{\mathbf{r},\delta=A,B}\left(\epsilon_{s}\hat{s}_{\mathbf{r},\delta}^{\dagger}\hat{s}_{\mathbf{r},\delta}+\sum_{\alpha}\epsilon_{t_{\alpha}}\hat{t}_{\mathbf{r},\delta,\alpha}^{\dagger}\hat{t}_{\mathbf{r},\delta,\alpha}\right),
\end{equation}
while the inter-dimer $H_{J}$ term are quartics of bond-operators
that provides off-diagonal quantum dynamics to the system.

The elementary quantum excitations above the dimer ground states can
be obtained by condensing the appropriate boson operators. Here we
first take the singlet dimer state as an example, and the same strategy
also applies to the triplet dimer states. In the singlet dimer state,
the singlet operator $\hat{s}$ is condensed
\begin{equation}
\hat{s}=\hat{s}^{\dagger}\approx\sqrt{1-\sum_{\alpha}\hat{t}_{\alpha}^{\dagger}\hat{t}_{\alpha}},
\end{equation}
while the triplet operators $\hat{t}_{\alpha}$ can be regarded as
elementary quantum excitations above the singlet dimer ground state.

Expanding the total Hamiltonian $H$ in terms of $\hat{t}_{\alpha}$
up to the quadratic order and performing the Fourier transformation

\begin{align}
\hat{s}_{\mathbf{r},\delta} & =\sqrt{\frac{2}{N}}\sum_{\mathbf{k}\in{\text{BZ}}}s_{\mathbf{k},\delta}e^{i\mathbf{R}_{\mathbf{r}}\cdot\mathbf{k}},\\
\hat{t}_{\mathbf{r},\delta,\alpha} & =\sqrt{\frac{2}{N}}\sum_{\mathbf{k}\in{\text{BZ}}}t_{\mathbf{k},\delta,\alpha}e^{i\mathbf{R}_{\mathbf{r}}\cdot\mathbf{k}},
\end{align}
at the quadratic order, we obtain the linear bond-operator Hamiltonian
in a compact matrix form
\begin{equation}
H=\frac{1}{2}\sum_{\mathbf{k}\in\text{BZ}}\Psi(\mathbf{k})^{\dagger}\mathbf{M(\mathbf{k})}\Psi(\mathbf{k})+const.
\end{equation}
where $\mathbf{r}$ denotes the position of the unit cell, $\delta=A,B$
represents the dimer index within the unit cell, $\alpha=1,2,3$,
$\text{BZ}$ denotes the Brillouin zone,
\begin{equation}
\Psi(\mathbf{k})=[t_{\mathbf{k},A,1},t_{\mathbf{k},A,2},...,t_{\mathbf{k},B,3},t_{\mathbf{-k},A,1}^{\dagger},t_{\mathbf{-k},A,2}^{\dagger},...,t_{\mathbf{-k},B,3}^{\dagger}]^{T},
\end{equation}
and $\mathbf{M(\mathbf{k})}$ is a ${12\times12}$ Hermitian matrix.
Then we can Bogoliubov diagonalize $H$ with ${\Psi(\mathbf{k})=T_{\mathbf{k}}\Phi(\mathbf{k})}$,
where
\begin{equation}
\Phi(\mathbf{k})=[\beta_{\mathbf{k},1},\beta_{\mathbf{k},2},...,\beta_{\mathbf{k},6},\beta_{\mathbf{-k},1}^{\dagger},\beta_{\mathbf{-k},2}^{\dagger},...,\beta_{-\mathbf{k},6}^{\dagger}]^{T}
\end{equation}
is the diagonalized basis of Bogoliubov quasi-particles, and $T_{\mathbf{k}}$
is the transformation matrix. The diagonalized Hamiltonian reads
\begin{eqnarray}
H & = & \frac{1}{2}\sum_{\mathbf{k}\in\text{BZ}}\Phi(\mathbf{k}){}^{\dagger}E(\mathbf{k})\Phi(\mathbf{k})+const.\nonumber \\
 & = & \sum_{\mathbf{k}\in\text{BZ}}\sum_{s=1}^{6}\omega_{\mathbf{k}s}\beta_{\mathbf{k}s}^{\dagger}\beta_{\mathbf{k}s}+const.,
\end{eqnarray}
where $E(\mathbf{k})=\textrm{diag}[\omega_{\mathbf{k}1},\omega_{\mathbf{k}2},...,\omega_{\mathbf{k}6},\omega_{-\mathbf{k}1},\omega_{-\mathbf{k}2},...,\omega_{-\mathbf{k}6}]$
is the diagonalized Bogoliubov Hamiltonian.

\subsect{Exact diagonalization}
Deep in the dimer phases, the dimers are effectively decoupled, therefore
it is save to ignore the inter-dimer interactions and consider independent
dimers. Under the basis $\psi\equiv\begin{pmatrix}|s\rangle & |t_{1}\rangle & |t_{2}\rangle & |t_{3}\rangle\end{pmatrix}^{T}$,
the Hamiltonian for each dimer can be represented by $4\times4$ matrices.
By diagonalizing such Hamiltonians, physical quantities such as energies
and magnetizations can be obtained.

\section*{Data availability}

The data that support the findings of this study are available from
the corresponding authors upon reasonable request.

\section*{Code availability}

The code that supports the findings of this study is available from
the corresponding authors upon reasonable request.

\section*{acknowledgments}

We thank A. Aczel, A. Brassington, J. Ma, 
X. Sun, and H. Zhou for useful discussions. This work
is supported by the National Key R$\&$D Program of China (Grant No.2023YFA1406500),
and the National Science Foundation of China (Grant Nos.12334008,
12174441 and 12564021).

\section*{Author Contributions}

The project was conceived by R.Y. and C.L.. Symmetry analysis was
performed by C.L.. Bond-operator calculation was performed by C.L.
and G.D.. Magnetization calculations were performed by G.D.. Theoretical
interpretation was provided by R.Y., C.L., G.D.. The manuscript was
written by C.L. and R.Y. with contributions from all the authors.

\section*{Competing interests}

The authors declare no competing interests.

%\bibliography{ref_new}

\end{document}